\documentclass[12pt]{iopart}

\usepackage[utf8]{inputenc}
\usepackage{natbib}
\usepackage{graphicx}
\expandafter\let\csname equation*\endcsname=\relax
\expandafter\let\csname endequation*\endcsname=\relax
\usepackage{amsmath,bm}
\usepackage{xcolor}
\usepackage{braket}
\usepackage{caption}
\usepackage{subcaption}
\usepackage{geometry}
\usepackage{float}
\usepackage{array}
\usepackage[normalem]{ulem}
\begin{document}

%\title{Efficient evaporative cooling down to 100 nK in a dipole trap in microgravity}

\title{Ultracold atoms in a dipole trap in microgravity}

\author{J. Le Mener$^1$ \& C. Métayer$^1$ \& V. Jarlaud$^2$ \& C. Pelluet$^3$ \&  B. Battelier$^1$}
%\author[1]{C. M\'{e}tayer}
%\author[1]{V. Jarlaud}
%\author[1]{C. Pelluet}
%\author[1]{J. Le Mener}
%\author[1]{P. Bouyer}
%\author[1]{B. Battelier}

\address{$^1$LP2N, Laboratoire Photonique Numérique et Nanosciences, Université de Bordeaux, IOGS and CNRS , 1 Rue François Mitterrand, 33400 Talence, France}

%\affil[1]{LP2N, Laboratoire Photonique Numérique et Nanosciences, Université de Bordeaux, IOGS and CNRS , 1 Rue François Mitterrand, 33400 Talence, France}

\address{$^2$Exail, 27 Avenue de la Poterie, 33170 Gradignan, France}

\address{$^3$Centre National d’Etudes Spatiales, 18 avenue Edouard Belin, 31400 Toulouse, France}

%\maketitle

\begin{abstract}
Most cold atoms experiments in microgravity platforms or in Space are achieved using atom chips, leading to limitations in terms of optical access and inhomogeneous magnetic fields. Optical dipole traps do not have these drawbacks but have difficulties producing atomic samples with a large number of atoms at ultra low temperature in the absence of gravity. 
Here, we report on an efficient evaporative cooling in two-crossed laser beams during parabolic flights. Time-averaged potentials combine the advantages of large capture volume and trap compression, increasing the initial phase space density and collision rate to favor the evaporative process. With this technique we demonstrate the production of an ultra cold gas of $2.5\times 10^4$ rubidium atoms at a temperature below 100 nK in less than 4 seconds. Our experiment paves the way for the development of quantum sensors applied to fundamental physics and geodesy as well as the study of ultracold atomic physics in Space. 
\end{abstract}
\vspace{10mm}

\section*{INTRODUCTION}
Cold atoms in microgravity is an emerging field which covers a large panel of physical phenomena, including, non exhaustively, few-body physics \cite{Naidon2017}, molecular spectroscopy and ultracold chemistry, exotic topologies and geometries \cite{Carollo2022,fryearndt2025}, and fundamental tests \cite{Battelier2021,Anton2025}.

A number of  Bose Einstein Condensates experiments have been performed on ground-based microgravity platforms \cite{vanZoest2010,Condon2019}  or in Space \cite{Becker2018,Aveline2020}, demonstrating the maturity and the robustness of the atom chip technology. Despite their promising performances, inherent difficulties due to the proximity of the atom chip are still to be tackled such as stray light, limited optical access, and inhomogeneous magnetic fields. %Solutions to transport the atoms far from the chip were demonstrated for a few mm only due to the decompression of the trap [Ref]. 
Alternatively, dipole traps have strong advantages to produce ultracold atoms, allowing high optical access, a fast extinction of the trapping potential and facilitating the control of the interactions with an independent control of magnetic fields for Feshbach resonances.

Unfortunately, the inherent decompression of the dipole trap prevents fast and efficient evaporative cooling, making  %while diminishing the trap depth through the laser power, thus 
the run-away regime unreachable \cite{OHara2001}.  
Unlike magnetic traps, the trap frequency decreases as the square root of the laser power while reducing the trap depth. On ground, the gravitational potential gradient decreases the potential barrier favoring the evaporation in the vertical direction and become significant during the final stage of the evaporative cooling. %(figure \ref{Fig: DipoleTrapPlane} (b)). 
In microgravity, the absence of this helping force leads to a new challenge. In return, the laser power can be further decreased compared to standard gravity \cite{Yuan2024} while maintaining the weak trapping force in the directions of the laser beams. %Practically, at low laser power, the effective trap depth is lowered in the direction of the beams, counter-balancing the absence of gravity. Consequently atoms are evaporated and potentially stay trapped in the weakly confining potential along the directions of the two beams $\hat{X}$ and $\hat{Y}$ (Fig \ref{Fig: DipoleTrapPlane} (c)). 
 %These particularities justify dedicated studies of production of ultracold atoms in a dipole trap in microgravity. 
Evaporative cooling \cite{Vogt2020} was demonstrated at high temperature ($>10 \mu$K) and a promising robust set up was tested in the HITec Einstein Elevator in Hannover \cite{Haase2025} but no Bose Einstein condensation in microgravity has been demonstrated so far.

\begin{figure}
    \centering
    \includegraphics[width=0.95\textwidth]{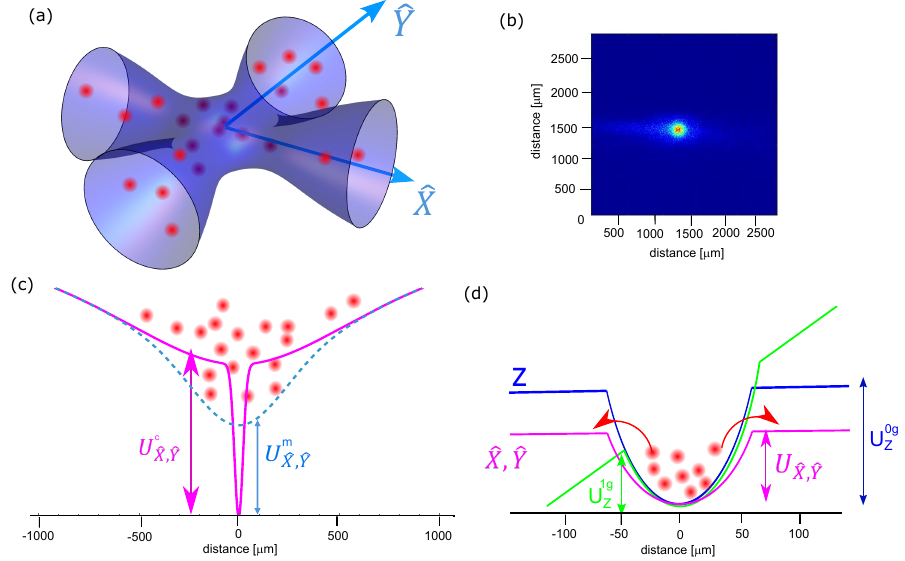}
    \caption{\textbf{Cooling and trapping in a dipole trap in microgravity:} (a) Atoms trapped in the two laser beams forming the crossed dipole trap. %play the role of a reservoir. 
    (b) Fluorescence imaging of the atoms in the dipole trap in microgravity. (c) Thanks to our painting potential, the trap is deformed at the crossing of the two beams and the increase of the trap depth from the modulated trap $U^m_{\hat{X},\hat{Y}}$ (blue) to the compressed trap $U^c_{\hat{X},\hat{Y}}$ (purple) leads to an increase in the phase space density before starting the evaporative cooling. (d) In standard gravity (green), the sag decreases the effective trap depth favoring the escape of the atoms in the vertical direction at the end of the evaporative cooling. In microgravity, % the laser power need to be lower and then leading to a decompression of the trap (dashed blue line). This effect is mitigated with the painted potential strategy (magenta), the small waist of the spatially modulated beam allowing a tight trap once the modulation is off.  
    there is no sag effect (blue) atoms are preferentially evaporated in the weakly confining potential along the directions of the two beams $\hat{X}$ and $\hat{Y}$ (purple).  }
    \label{Fig: DipoleTrapPlane}
\end{figure}

In this work, we demonstrate the production of ultracold atoms in a dipole trap in microgravity and reach a phase space density at the threshold of Bose Einstein condensation. Our painted potential created by spatial modulation of a tightly focused laser beam allows for a good trade-off between capture volume and trap depth  \cite{Condon2019}. Due to the influence of the laser beams, the atoms are trapped in a bimodal potential. The spatial modulation is turned off to change the shape of the trap slowly compared to the internal
equilibration time \cite{Pinkse1997}, increasing the phase space density. In the meantime the compression of the trap increases the collision rate favoring efficient evaporative cooling. 
%In the following, we present our experimental sequence to load and cool the atoms in the dipole trap. We characterize the efficiency of the evaporative cooling and compare it to the case of standard gravity. 
We demonstrate that it is possible to reach high phase space density despite a decrease in collision rate due to the decompression of the trap. %The experiment was produced on the 0-g plane performing parabolic flights (Fig. \ref{Fig: DipoleTrapPlane} (a)). 
Time of flight measurements with a duration of free fall up to 100 ms allow us to evaluate the spatial expansion of the ultracold atomic sample.

\section*{Results and discussion}

The results were obtained onboard the Novespace 0g aircraft performing parabolic flights. % We previously demonstrated that we are able to produce Bose-Einstein condensates (BEC) on an Einstein elevator in the laboratory. 
 %A major difference with the experiment presented here is that the entire sequence takes place in microgravity whereas on the Einstein elevator, evaporative cooling was performed during successive regimes of weight ($1g-2g-0g$). 
A three-dimensional magneto-optical trap (MOT) is loaded in the vacuum chamber from a two-dimensional MOT during $1$ s, leading to $1.5\times 10^{8}$ trapped rubidium atoms. Then the atom cloud is cooled down to $4.5\mu$K in $9$ ms using red optical molasses. After the laser cooling stage, our method \cite{Condon2019} combines simultaneously grey molasses cooling and trapping in a time-averaged optical potential. An amplified fibered telecom laser delivers up to 23W of $1550$ nm light and the dipole trap is formed by two crossed beams (Fig. \ref{Fig: DipoleTrapPlane} (a)) spatially modulated by an acousto-optic modulator (AOM) to increase the capture volume. Approximately $6\times10^{6}$ atoms are loaded in the dipole trap in $150$ ms (Fig. \ref{Fig: DipoleTrapPlane} (b)).  The variation in the acceleration between the steady flight, the hypergravity and microgravity phase of the parabola leads to a relative misalignment of the two crossed beams of the dipole trap. To mitigate this effect, a real time compensation system realigns the relative position of the two beams at each experimental sequence (see Fig. \ref{DipoleTrapSetUp} in Methods). 

In the directions  $\hat{X},\hat{Y}$ of the two laser beams propagation, the dipole trap potential has a double structure with an additional trapping force due to the focus of the propagating beam. %corresponding to a residual trap frequency $\omega_{res}$.  
This effect is significant here because the painted potential allows the beam to be focused strongly on the atoms without reducing the capture volume defined by the spatial modulation amplitude.
Atoms are thus trapped in a bimodal potential which plays an important role in increasing the initial space density  (Fig. \ref{Fig: DipoleTrapPlane} (c)). %The atoms trapped in the laser beams play the role of a reservoir for the high density atomic cloud at the center of the crossed trap. Additionally The gain in phase space density
which can be understood with a simplified two-box model, $V_0$ being the volume in the laser beam and $V_1<<V_0$ the volume in the crossed dipole trap  \cite{StamperKurn1998}:

\begin{equation}
\ln{\left(\frac{D_1}{D_0}\right)}=\frac{U_1/k_B T}{1+(V_1/V_0)e^{U_1/k_B T}}
\label{eq:GainPSDBox}    
\end{equation}
$U_1$ is the potential energy in the small box, where the increase in atomic density can lead favorably to a local gain in phase space density $D_1>D_0$. %In the non modulated case, the trap frequency in the crossed trap is much higher than $\omega_{res}$, leading to $\omega_{\hat{X},\hat{Y}}/\omega_{res}= \frac{\sqrt{2} \pi w_0}{\lambda} \approx 120 $ for a waist $w_0=42.5 \mu$m.

\begin{figure}
    \centering
    \includegraphics[scale=0.5, angle=0]{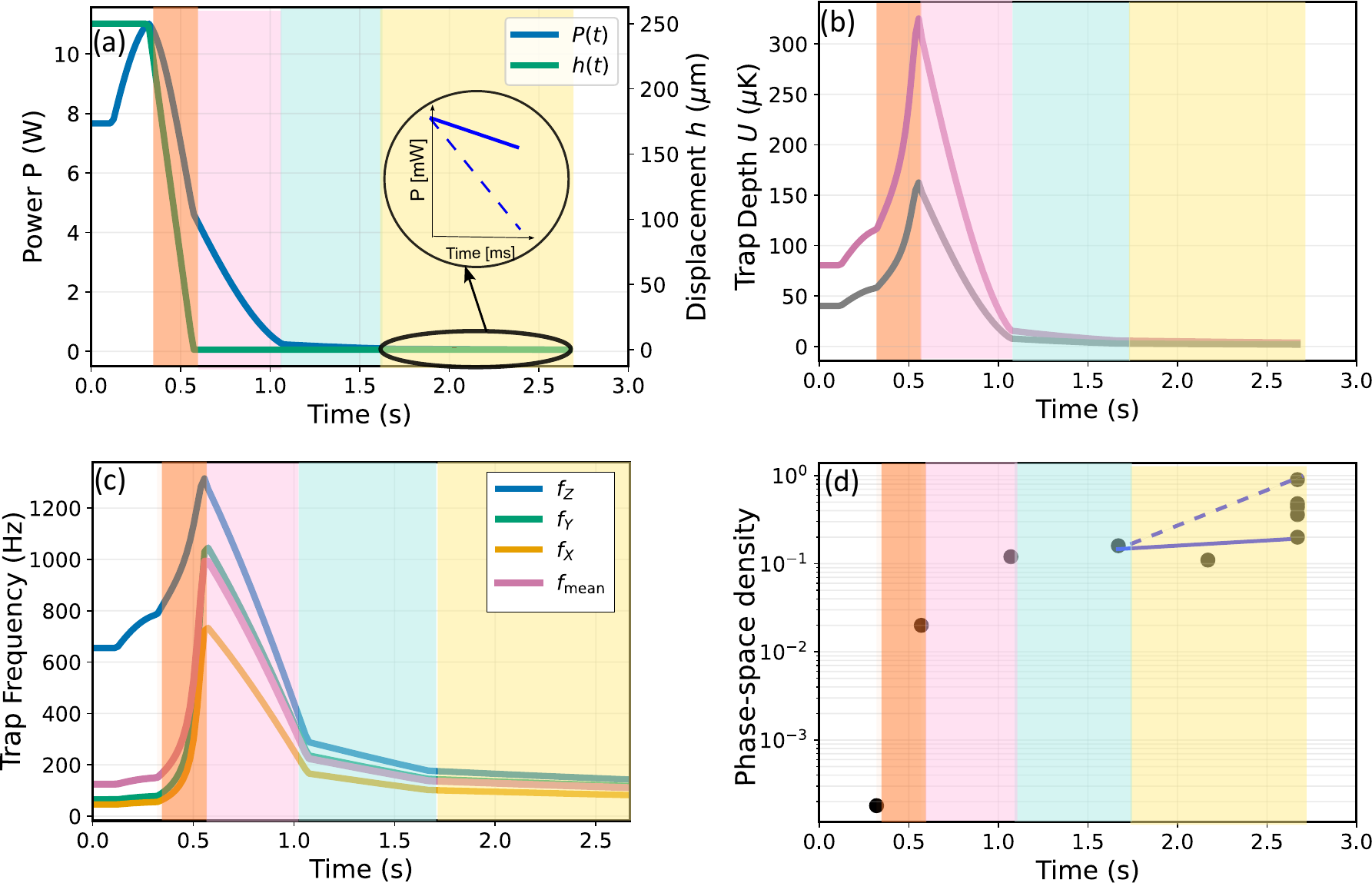}
    \caption{\textbf{Production of ultracold atoms in a crossed dipole trap in microgravity:} (a) Laser power (blue) and spatial modulation amplitude $h$ (green) during the compression phase (orange zone) and the three phases of evaporative cooling (pink, light blue and yellow respectively). Inset: To improve the evaporative cooling in microgravity, we decrease the final value of the laser power (dashed blue line) at the end of the final ramp. The solid line is the reference ramp in standard gravity. (b) Trap depth in the Z direction (purple) and in the direction of the two crossed beams $\hat{X},\hat{Y}$ (khaki). %during the loading, compression and evaporative cooling. %\textcolor{red}{Attention pas un facteur 2 quand le piège est modulé spatialement. A faire avec la simulation de Julien} 
    (c) Trap frequencies along the eigen-axes of the dipole trap ($X,Y,Z$) and (d) phase space density in the dipole trap in microgravity (blue).  We measured a gain in the phase space density of 2 orders of magnitude during the compression phase (orange). Reducing the final value of the last ramp (dashed line) leads to a final gain in the phase space density.}
    \label{fig: Evap Picture}
\end{figure}

The painting potential allows deforming the shape of the center part of the trap adiabatically \cite{Pinkse1997}. The spatial modulation is ramped down to zero in $\tau_c=250$ ms (Fig. \ref{fig: Evap Picture} (a)) and leads to an increase in the crossed trap depth $U_c \approx 3 U_m$ (Fig. \ref{fig: Evap Picture} (b)), where the index $c$ means compressed trap and $m$ stands for the modulated trap. The trap depth rise from $U_m$ to $U_c$ during the compression phase leads to an increase in phase space density in the crossed dipole trap $D_c>D_m$:

\begin{equation}
\ln{\left(\frac{D_c}{D_m}\right)}=\frac{U_c/k_B T_c}{1+(V_c/V_0)e^{U_c/k_B T_c}}- \frac{U_m/k_B T_m}{1+(V_m/V_0)e^{U_m/k_B T_m}}
\label{eq:GainPSD}    
\end{equation}
with $V_{c,m}$ the volume in our crossed dipole trap without and with spatial modulation respectively.
We measure $T_c\approx T_m \approx 13 \mu$K which is consistent with an adiabatic compression of the trap $\tau_{c}\approx 250\delta\bar{\omega}/\bar{\omega}^2$, $\bar{\omega}$ being the mean trap frequency. The volume along the two laser beams being much larger than the volume in the crossed trap  $V_0>>V_{m,c}$, Eq. \ref{eq:GainPSD} leads to a gain in phase space density proportional to the Boltzmann factor  $D_c/D_m\approx\exp{[(U_c-U_m)/k_B T_c}]$. We measured a gain of two orders of magnitude experimentally corresponding to $(U_c-U_m)/(k_B T_c)\approx 5$ which is reached during the compression phase. After this value, the two-box model is not valid anymore because most of the gas become confined in the crossed dipole trap and then the evaporative cooling starts.
%\textcolor{red}{bémols: l'effet s'arrete avant Uc max à 160 microK, ca colle pour 120 microK, mais pourquoi pas  on l'atteint en cours de route, en suite trop profond } %Thus, the phase space density at the center of the trap of about two orders of magnitude to reach $ 2\times 10^{-2}$, which is consistent with .

During the compression phase, the trap frequencies (Fig. \ref{fig: Evap Picture} (c)) and therefore the atomic density are increased, leading to favorable conditions for evaporative cooling which is performed by progressively lowering the power of the dipole trap laser following three linear ramps (Fig. \ref{fig: Evap Picture}(a)) to decrease the trap depth (Fig. \ref{fig: Evap Picture}(b)). 
 The phase space density increases by one order of magnitude during the first linear ramp (pink on Fig. \ref{fig: Evap Picture} (d)), and then stays constant until the final stage of the evaporative cooling sequence. During this last stage at low laser power, there is no gravity sag supporting evaporative cooling in the vertical direction as in standard gravity (Fig \ref{Fig: DipoleTrapPlane} (d)).  
 To force the evaporation process, the final value of the laser power is further lowered (Fig. \ref{fig: Evap Picture} (a) inset), leading to a final increase in the phase space density (Fig. \ref{fig: Evap Picture} (d)). %This effect is similar in standard gravity and microgravity (see figure \ref{fig: Evap Picture} (d)). Indeed, the gravity potential impacts the trap depth at the level of $2.5 \mu$K, which has no significant impact during the loading and compression phases where $U_{\hat{X},\hat{Y}}>75 \mu$K ((Fig. \ref{fig: Evap Picture}(b)) is much larger.

\begin{figure}
    \centering
    \includegraphics[scale=0.37, angle=0]{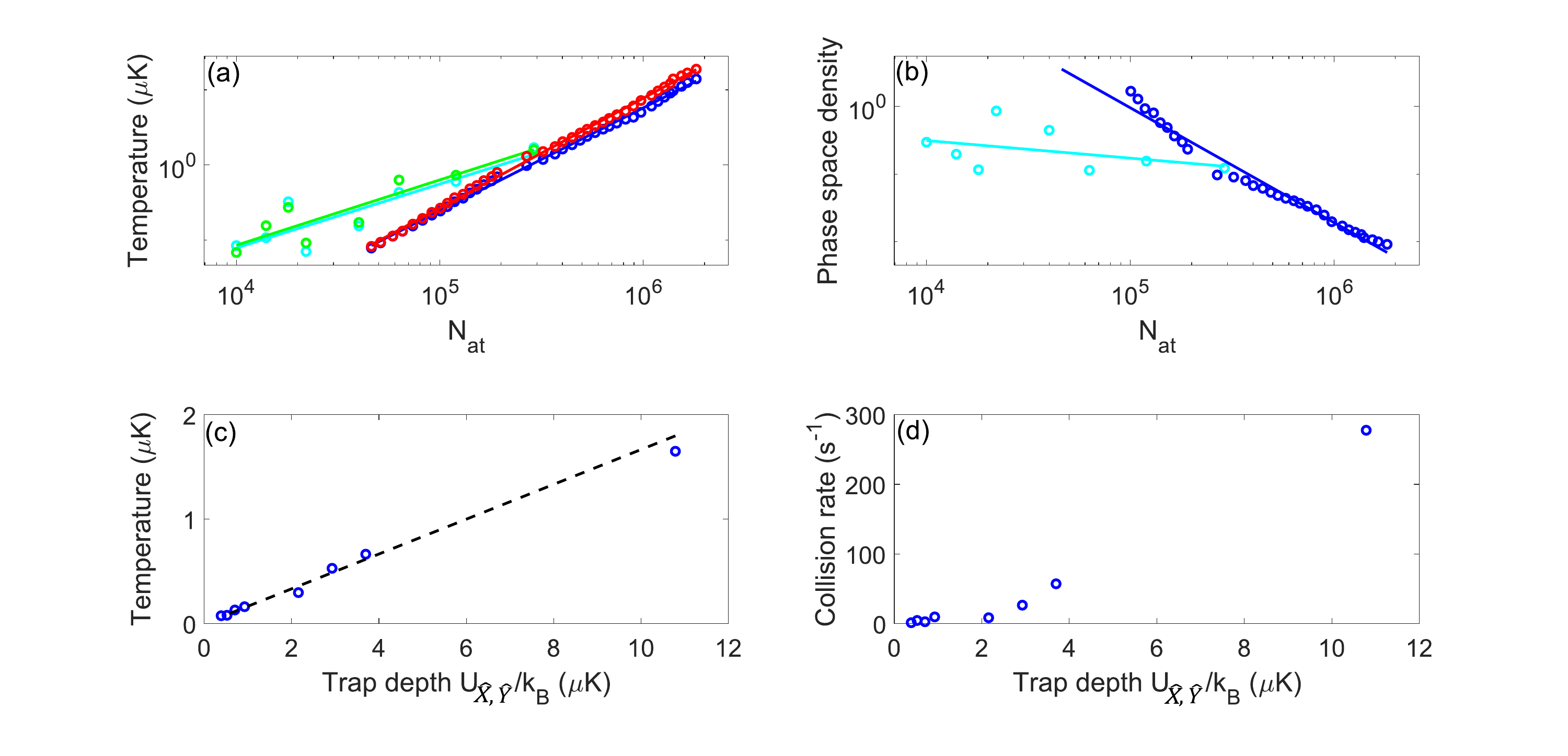}
    \caption{\textbf{Parameters of the evaporative cooling in the dipole trap in microgravity:} Temperature (a) and Phase Space Density (b) versus the atom number in microgravity onboard the 0g aircraft (green and light blue) and in standard gravity (blue and red). The slope allows us to extract the parameters $\alpha_T$ and $\alpha_D$ giving quantitative estimation of the cooling efficiency. (c) Temperature and (d) collision rate versus the trap depth. The data shows $\eta=U_{\hat{X},\hat{Y}}/(k_B T)\approx 6$ during the evaporative cooling (dashed line on (c)). } 
    \label{fig:EvaporativeCooling}
\end{figure}

We compare the efficiency of the evaporative cooling in our apparatus respectively in microgravity and standard gravity (Fig. \ref{fig:EvaporativeCooling}). Starting from the initial temperature $T_{ini}$, atom number $N_{ini}$ and phase space density $D_{ini}$, the parameters $\alpha_T$ and $\alpha_{D}$ link respectively the temperature and the phase space density with the atom number: 

\begin{equation}
\frac{T_{fin}}{T_{ini}}=\left(\frac{N_{fin}}{N_{ini}}\right)^{\alpha_T}    \label{eq:AlphaEvap}
\end{equation}

and 

\begin{equation}
\frac{D_{fin}}{D_{ini}}=\left(\frac{N_{fin}}{N_{ini}}\right)^{\alpha_D}    \label{eq:AlphaEvap}
\end{equation}

These scaling laws for evaporative cooling in an adiabadically decompressed and lowered optical trap depend on the trap depth $\eta=U_{\hat{X},\hat{Y}}/(k_B T)$ which is lower along the $\hat{X}$ and $\hat{Y}$ direction because the potential energy is non-zero inside the laser beams and the most energetic atoms in the crossed trap preferentially escape along the beam propagation. Consequently $\eta\approx 6$ gives respectively $\alpha_T=2 (\eta+\kappa-3)/3$ and $\alpha_D=4-(\eta+\kappa)$, $\kappa$ being a constant number close to 1 \cite{OHara2001}.
Fig. \ref{fig:EvaporativeCooling} (a) shows the temperature versus the atom number and allows us to extract $\alpha_T$.  
In microgravity, the potential due to the beam in the propagation direction lowers the effective trap depth $\eta$ leading to effective evaporative cooling in microgravity with $\alpha_T^{0g}\approx 0.9$ which is lower than in standard gravity where $\alpha_T^{1g}\approx 1.4$.

Similarly Fig. \ref{fig:EvaporativeCooling} (b) presents the phase space density versus the atom number during the evaporative process and allows us to extract $\alpha_D$. In standard gravity, the evaporative cooling leads to a continuous increase in the phase space density until Bose-Einstein condensation with $\alpha_D^{1g}\approx -1.7$. In microgravity $\alpha_D^{0g}\approx -0.5$ is much lower than the expected value $\approx -3$ for $\eta\approx 6$ which indicates significant losses with a characteristic loss rate $\Gamma_{\rm loss}^{-1} \approx 640$ms due to vibrations and relative misalignements of the beams. By comparison these losses prevail over residual heating limiting the lifetime in the dipole trap which is evaluated around $ 18$ s on ground.  %Since there is no gravity sag, it is necessary to decrease the trap depth at a lower value compared to standard gravity. 
Fig. \ref{fig:EvaporativeCooling} (d) shows the evolution of the collision rate which decreases significantly at the end of the evaporative cooling ($\gamma= 6$ s$^{-1}$) preventing us from shortening the duration of the last power ramp to mitigate the losses. The data presented in Fig. \ref{fig:EvaporativeCooling} are acquired over the 3 flights of the campaign. The conditions can be slightly different from one day to another, which can explain the fluctuations of the phase space density in the characterisation curves.
%$\alpha=2(\eta+\kappa-3)/3$ link between the two parameters (Based on Claude Cohen Tanoudji College de France 1996-1997). The factor 2 in $\alpha$ compared to magnetic traps takes into account the decompression of the dipole trap. 

  %\textcolor{red}{Temperature plus grande que la profondeur du piège (transverse au faisceau), mais OK pour la direction du faisceau, le laser "guide"  grossierement les atomes qui sévaporent}

%\begin{figure}
%    \centering
%    \includegraphics[scale=1.0, angle=0]{Fig PSDmicrogravity v2.pdf}
%    \caption{\textbf{Increase of Phase Space Density in microgravity:} \textcolor{red}{Redondance PSD avec la figure 3, meme si pas tracé pareil} Trap frequencies along the eigen axes of the trap during the compression phase and the evaporative cooling \textcolor{red}{PLEASE USE Z as the vertical axis!!!}. Phase Space Density during evaporative cooling in microgravity (blue) and in standard gravity (orange). The grey zone corresponds to a gain of the PSD of 2 orders of magnitude due to the compression of the trap.}
%    \label{fig:PSDEvaporative}
%\end{figure}

%\begin{itemize}
%\item PSD$\approx 2.6$ \textcolor{red}{Why 1 in Clement's thesis?} 
%\item peak atomic density: $~9 \times 10^{12}$ cm$^{-3}$ (typically $10^{14}$ at.cm$^{-3}$ pour un condensat} 
%\item Critical temperature $T_c=100$ nK (N=25000 \textcolor{red}{TBC} without the atoms in the laser beams)
%\item Mean Trap frequencies: 79 Hz
%\item Collision Rate $\gamma= 15$ s$^{-1}$
%\end{itemize}

\begin{figure}
    \centering
    \includegraphics[scale=0.35, angle=0]{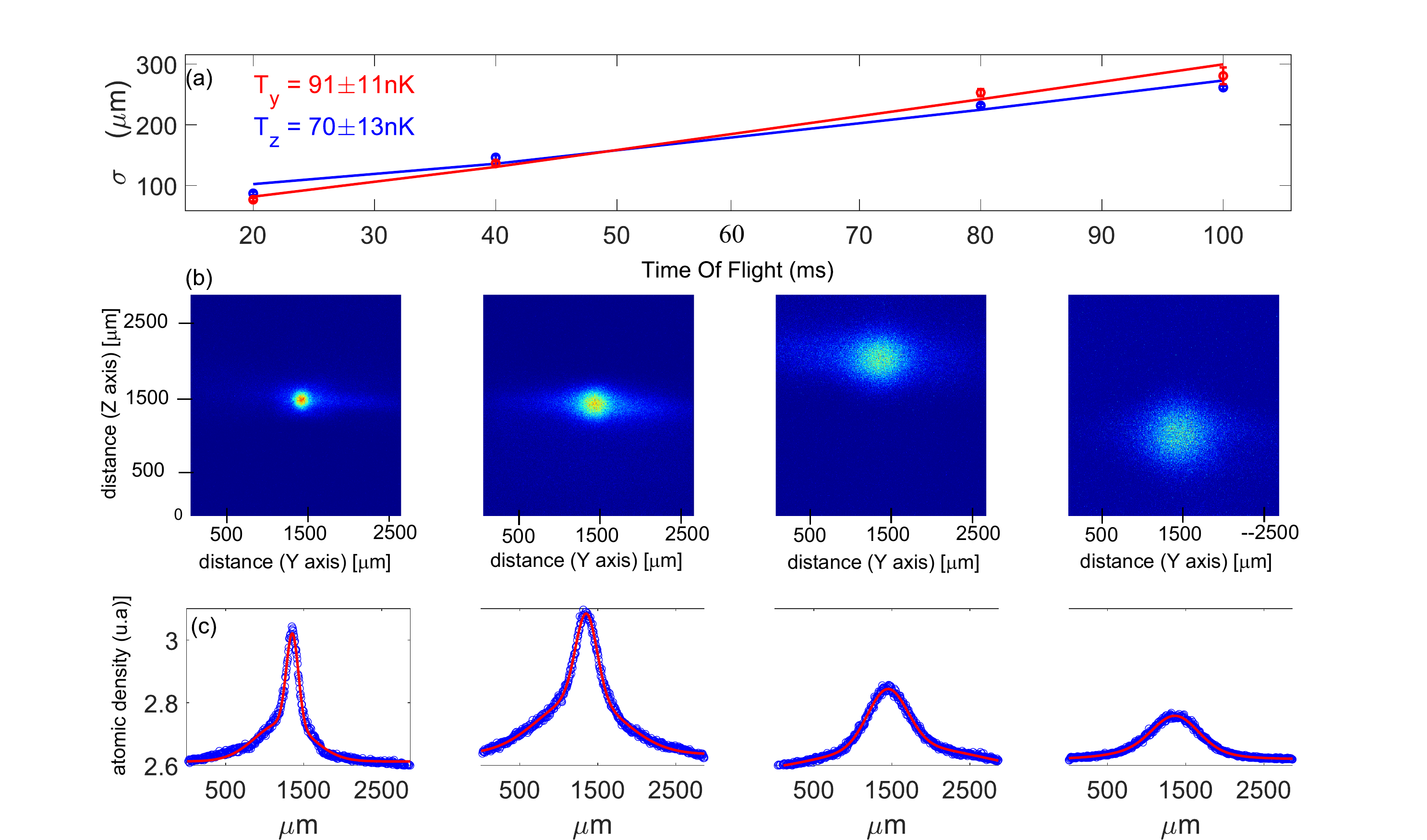}
    \caption{\textbf{Ultracold atoms in microgravity:} %\textcolor{red}{CAMPAGNE DE VOLS 2024: Vol 3, parabole 4 a 7 PSD:2.9} 
    (a) Spatial expansion of the atom cloud in the Y and Z direction %$\sigma_Y$ and $\sigma_Z$ 
    versus the time of flight. The slope gives the equivalent temperature of the atomic sample. (b) Fluorescence detection of the spatial profile of the atom cloud for four durations of free fall from left to right: 20 ms, 40 ms, 80 ms and 100 ms. (c) The atom cloud profile is integrated along the Z direction and fitted along Y by a double Gaussian profile (the experimental data is in blue and the fitted profile in red).}
    \label{fig:TimeOfFlight}
\end{figure}

A single parabola of 20 seconds of microgravity allows us to perform a full temperature estimation by varying the time of flight of the atoms released in free fall (4 or 5 measurements). At the end of evaporative cooling, the dipole trap is turned off and the radio-frequency sent to the AOM is detuned by 2 MHz to change the position of the dipole trap by 300 $\mu$m to make sure that no residual light leads to any force on the atoms.
By measuring the expansion of the ultracold gas in free fall using fluorescence imaging, we find a temperature of approximately $80$ nK (Fig. \ref{fig:TimeOfFlight}). The maximum free fall duration is limited to $100$ ms because the atomic sample moved out of the detection zone of the EMCCD camera ($3 \times 3$ mm$^2$ in the object plane) following the residual accelerations of the aircraft. This problem could be mitigated by choosing an imaging system with a lower magnification.

As a result, the process leads to a phase space density $D_{fin}\approx0.9$ at the center of the atomic cloud, close to the Bose Einstein condensation expected for $D_{\rm crit}=g_{3/2}(1)\approx 2.612$, corresponding to the critical temperature $T_c=100$ nK for $N=25000$ atoms and a mean trap frequency $\bar{\omega} =2\pi \times 59$ Hz. Additionally, the trap frequency along the laser beam propagation is weak $\omega_{beam}\approx 0.5 $Hz but strong enough to maintain atoms in microgravity (Fig. \ref{fig:TimeOfFlight} (b) and (c)). %, at ultracold temperature below 100 nK. 
To take this effect into account, the images of the atoms located in both the crossed trap and the laser beams are fitted with a double Gaussian profile. We thus estimate the atom number trapped inside the two laser beams $N_{\rm beam}\approx 17000$ for a temperature of about 1.5 $\mu$K.

In conclusion, we successfully demonstrated the production of ultracold samples of 25 000 atoms at 80 nK in a dipole trap in microgravity, reaching the threshold of quantum degeneracy in less than 4 seconds. To our knowledge, these results correspond to the lowest temperature achieved with an all-optical method entirely performed in microgravity.  Although cooling efficiency is reduced in the absence of gravity, the use of the painted potential strategy has made this achievement possible by focusing the laser beam to a small waist while maintaining a large capture volume. The compression phase not only ensures a high initial collision rate allowing evaporative cooling but also a significant increase in the phase space density due to the adiabatic deformation of the bimodal potential in the direction of the laser beams. We are confident that further improvements, such as reducing the waist or increasing the initial atom number with 2D spatial modulation, will lead to the production of Bose Einstein condensation in microgravity.

%Pursuing our efforts is crucial to understand the underlying mechanisms in microgravity 
Compared to other microgravity platforms, the easy access of the 0g aircraft leads to a good control of the experimental parameters and makes our experiment a unique testbed to prepare future space missions in fundamental physics, geodesy and navigation. %In flight we have been able to adapt the laser power variation to optimise "in real time" the PSD. Unlike the standard gravity, a faster decrease in laser power at the end of evaporative cooling is required due to the reduced collision rate caused by trap decompression. While this approach shows great promise, we anticipate further optimizations through a more detailed study in microgravity. 
The main limitations are related to the residual acceleration of the aircraft and are expected to be overcome in cleaner microgravity environments such as drop towers \cite{Muntinga2013,Raudonis2023,Lotz2017}, space Stations \cite{Aveline2020, He2023} or a satellite \cite{Leveque2021}.  

Dipole traps in microgravity will play a key role to guarantee the overlap of two species atomic sources which is critical to test the Weak Equivalence Principle \cite{Battelier2021}. More generally in quantum gases physics, such optical trap allows controlling the atom-atom interactions with independent magnetic fields \cite{Naidon2017}. Moreover, recent proposals are based on the optical dipole force to produce new trap geometries such as shells \cite{veyron2025,Wolf2022,Jia2022}, or boxes \cite{Navon2021,fryearndt2025} paving the way towards new topological phenomena.

\section*{METHODS}

\section*{Automatic realignment of the two laser beams of the dipole trap}

% A differential pumping stage guarantees  good ultra high vacuum level in the science chamber with a measured lifetime in the dipole trap longer than 10 seconds.

\begin{figure}
    \centering
    \includegraphics[width=0.95\textwidth]{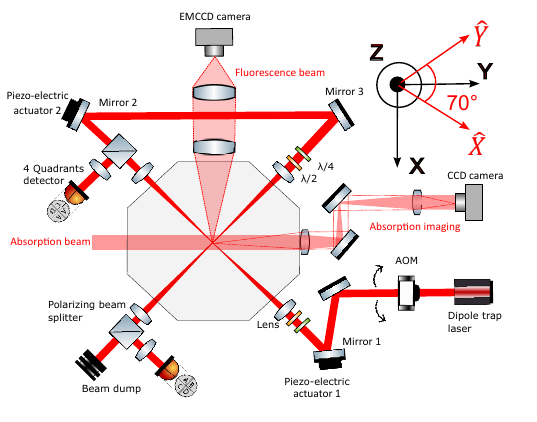}
    \caption{\textbf{Time averaged crossed dipole trap in microgravity:} schematic of the apparatus including the two crossed beams of the dipole trap, two imaging systems (absorption and fluorescence respectively) and the automatic realignment systems of the dipole trap laser beams. The positions of each beam are measured with a 4 quadrant detector and the feedback is applied to the mirror 1 and 2 respectively using piezo electric actuators. In microgravity, the atoms stay at the center of the vaccuum chamber. AOM: Acousto-optic modulator. $\lambda/2$ :halfwave plate. $\lambda/4$ :quarterwave plate. %The detection regions of the imaging systems are centred on the position of the dipole trap which also corresponds to the point of convergence of the cooling lasers. 
    }
    \label{DipoleTrapSetUp}
\end{figure}

Fig. \ref{DipoleTrapSetUp} (a) gives a schematic representation of the science chamber, including the vacuum system and the dipole trap set up. The laser beam of the dipole trap goes through an acousto-optic modulator (AA Opto-electronic MTS40-A3-1550), used to control the power and modulate spatially the position of the beam. A pair of mirrors reflects the beam back to the vacuum chamber after a first pass in order to form the crossed dipole trap with an angle of 70°.  The two laser beams are focused at the centre of the vacuum chamber. The distance between the two mirrors M1 and M2 is $420$mm and are positioned at a distance which is approximately $370$ mm from the centre of the vacuum chamber. 

\begin{figure}
    \centering
    \includegraphics[scale=0.4, angle=0]{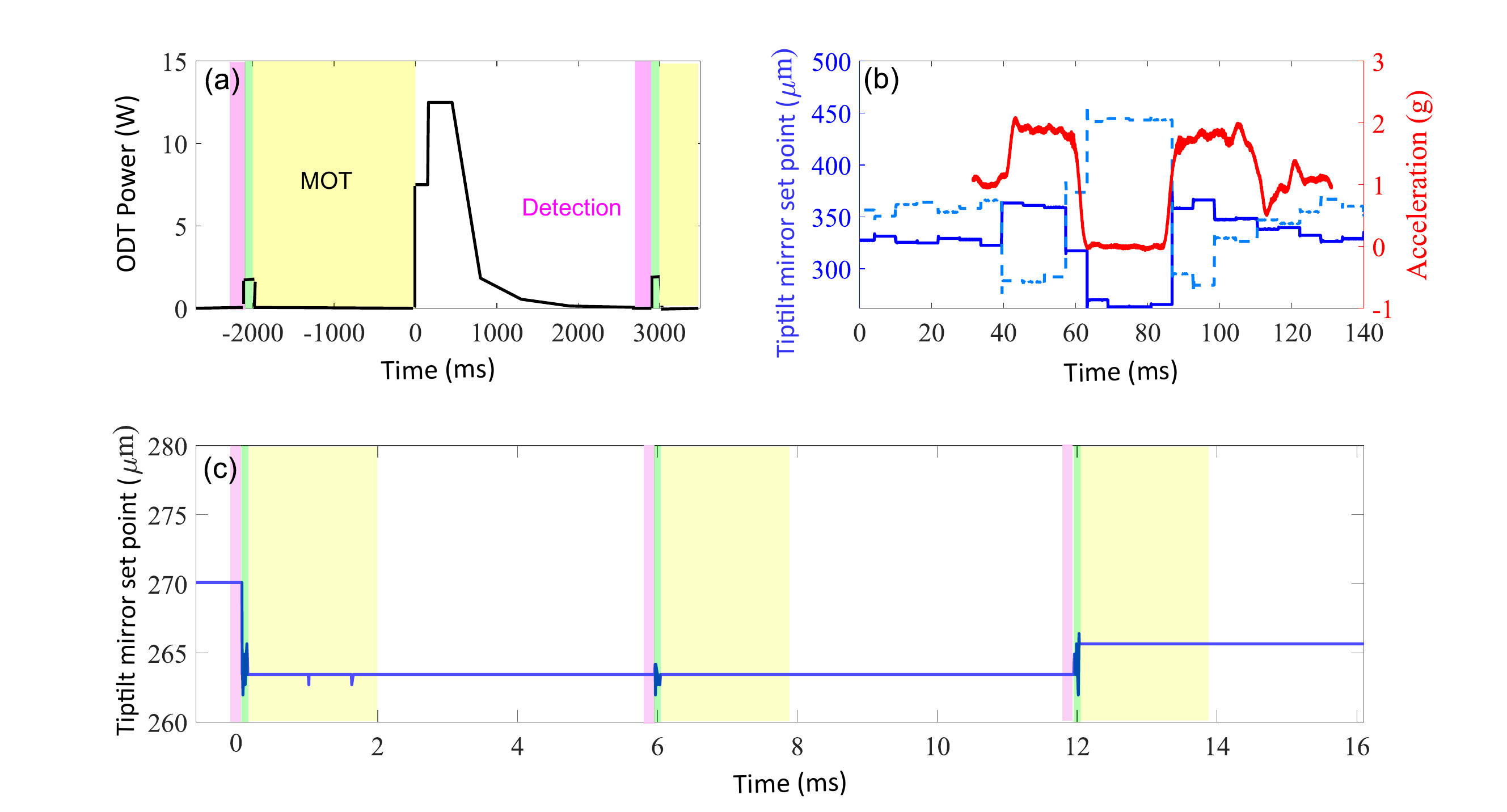}
    \caption{\textbf{Automatic realignment of the two dipole trap beams}: (a) Timing of the experimental sequence. The servo locks of the beam positions are ON during 100 ms (green) between the detection (pink) and the next MOT phase (yellow). (b) Acceleration (red) and Feedback signal on the piezo-electric mirror (dark and light blue) to maintain the two laser beams on the position detectors during a full parabola. (c) Principle of the "sample and hold" procedure to realign the two beams of the dipole trap over 3 experiment cycles. }
    \label{fig:PositionLockTrapBeam}
\end{figure}

To correct in real time the relative misalignment of the two crossed beams, the relative position of each beam is detected on an InGaAs 4-quadrant detector, by sampling $1\%$ of the laser power with a beamsplitter. Servo-lock electronics send feedback signals to piezoactuators which control the tilts of the two mirrors M1 and M2. Since the 4-quadrant detector does not have enough dynamic range in power to cover the full sequence of evaporative cooling, we apply a sample and hold technique. A constant laser power is applied to realign the two beams during the sample phase just before the MOT during 100 ms (Fig. \ref{fig:PositionLockTrapBeam} (a)). The strong acceleration variations during the successive phases of standard, hyper- and microgravity lead to significant misalignments of the two laser beams. The compensation system applies a correction of the tilts of the two mirrors M1 and M2 to automatically realign the beams without the intervention of the experimenter (Fig. \ref{fig:PositionLockTrapBeam} (b)). Inside the 0g phase, the residual misalignments of the beam $\approx 5-10 \mu$m are smaller than the beam waist, allowing the production of ultracold atoms (Fig. \ref{fig:PositionLockTrapBeam} (c)). 

\section*{Detection calibration}

\begin{figure}
    \centering
    \includegraphics[scale=1.8, angle=0]{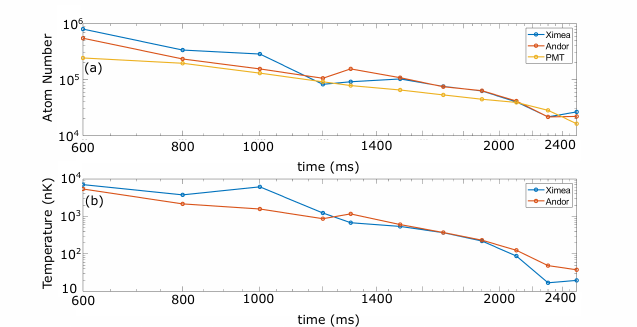}
    \caption{\textbf{Detection calibration using three methods}: absorption imaging (blue), fluorescence imaging (red) and fluorescence detection using a photomultiplier tube PMT (yellow). (a) Atom number and (b) temperature of the atom cloud during the evaporative cooling in standard gravity.}
    \label{fig:Detection}
\end{figure}

Three detection systems are installed on our set up (Fig. \ref{DipoleTrapSetUp}). The experiments in microgravity are done with the fluorescence imaging device. The two other detection systems, an absorption imaging set up and a fluorescence detection using a photomultiplier tube (PMT), are used for a calibration on ground, comforting us in the evaluation of the atom number and the temperature measured by spatial expansion during the free fall of the atom cloud. We compared the methods during a sequence of evaporative cooling (Fig \ref{fig:Detection}). 

For both fluorescence and absorption imaging, the temperature $\mathcal{T}$ is estimated by varying the time of free fall $t_{tof}$ after turning the dipole trap off and is extracted from the width of the cloud fitted with a Gaussian profile:
\begin{equation}
\sigma_X=\sqrt{\sigma_{X,0}^2+\frac{k_B \mathcal{T}}{m}t_{tof}^2}
\label{eq:TemperatureToF}
\end{equation}
where $\sigma_{X,0}$ is the initial size before the free fall.

We detail below the three methods to evaluate the atom number.

\subsection*{Fluorescence imaging detection}

Atoms are detected by fluorescence imaging with an electron multiplying CCD (EMCCD) camera which is adapted for low atomic density (Fig. \ref{DipoleTrapSetUp}). The fluorescence imaging beam propagates along the axis ($z$) perpendicular to the dipole trap plane, is circularly polarised and not retro-reflected. The photons emitted by the atoms are collected thanks to a 2 inch diameter lens of focal length  $f=150$ mm associated with a lens of focal length $f=400$ mm to image the atomic cloud on the EMCCD camera with a magnification of 2.7.\\

%While operating on ground, three detection systems are available to characterize and compare them: a photomultiplier (PMT), a fluorescence spatial imaging and the absorption imaging. 

%\subsection*{Spatial fluorescence imaging}

The diffusion rate of the photons emitted by the atoms depends on the intensity $I$ and the detuning $\Delta$ of the laser beam compared to the D2 line transition :

\begin{equation}
    \gamma_{diff} = \frac{\Gamma}{2} \cdot \frac{\frac{I}{I_{sat}}}{ 1 + 4\cdot (\frac{\Delta}{\Gamma})^2 + \frac{I}{I_{sat}}}  \label{eq:diffusionRate}
\end{equation}

with $\Gamma = 2 \pi \cdot 6.066$~MHz the natural linewidth of rubidium, and $I_{sat}$ the saturated absorption. The total photon number collected by the optical system and the camera during the exposition time $t_{exp}$ is directly related to the atom number $N_{at}$:

%\begin{equation}
%    N_{p,tot} = 
%\end{equation}
\begin{equation}
    N_p = \frac{\Omega}{4 \pi}N_{at} \gamma_{diff} t_{exp}
\end{equation}

where $\Omega$ is the solid angle of the optical system. The atom cloud measured on the EM-CCD camera is fitted with a Gaussian profile. The integration of the 2D profile leads to the total number of counts $N_c$, leading to the total atom number:

\begin{equation}
    N_{at}  =  \frac{4 \pi}{\Omega} \frac{N_c \xi}{\gamma_{diff} t_{exp} \eta G} 
\end{equation}

where $\eta$ is the quantum efficiency, $G$ the Electron Multiplying (EM) gain and $\xi$ the sensitivity of the camera in electron per count. The parameters are provided by the datasheet of the EMCCD camera and are confirmed by our detection calibration by comparison with the other detection methods.

\subsection*{Absorption imaging}

Atoms are detected by absorption imaging with a CCD camera Ximea MD120MU-SY. The optical system is composed of a first lens with a focal length f= 150 mm associated to a second lens of focal length 125 mm leading to a magnification of 0.83. Optical Density (OD) is extracted from a series of three images:  

\begin{equation}
OD (x,z)=\left(1+4\frac{\Delta^2}{\Gamma^2}\right)\ln\left(\frac{I_{beam}(x,z)}{I_{at}(x,z)}\right)%+\frac{I_{beam} (x,z)-I_{at} (x,z)}{I_{sat}}
\end{equation}
with $I_{at}(x,z)$ and $I_{beam}(x,z)$ corresponding to the two images taken with and without atoms respectively, subtracted by a third image measuring the background without light. The exposure time of each image is 70 $\mu$s, separated by a delay of 70 ms.

The total atom number $N_{at}$ is estimated from the widths $\sigma_x$ and $\sigma_z$ of the optical density $OD (x,z)$ fitted with a 2D Gaussian profile and the maximum optical density $OD_{max}$:

\begin{equation}
    N_{at} = \frac{2 \pi \sigma_x \sigma_z}{\sigma} \cdot OD_{max}
\end{equation}

%avec une gaussienne definit par 
%\begin{equation}
%    G(x,y) = OD\cdot e^{\frac{-x^2}{2 \sigma_x^2}}e^{\frac{-y^2}{2 \sigma_y^2}} + OD_0
%    \label{eq:gauss2D}
%\end{equation}

$\sigma$ is the scattering cross section depending on the intensity $I$ and the detuning $\Delta$ of the laser beam: 

\begin{equation}
    \sigma = \frac{\sigma_0}{ 1 + 4 \left( \frac{\Delta}{\Gamma} \right)^2 + \frac{I}{I_{sat}} }  
\end{equation}

where  $\sigma_{0}=1.938 \cdot 10^{-13}$ m$^{-2}$ is the on- resonance cross section of Rubidium 87 at resonance and for a linear polarisation \cite{STECK}. Experimentally the intensity of the absorption beam is set up to $I \approx 0.55$~mW.cm$^{2}$ and is at resonance $\Delta=0$ leading to a scattering cross section $\sigma \approx 5/6\cdot\sigma_0$.

\subsection*{Temporal fluorescence detection}

The fluorescence emitted by the atom cloud is measured by a photomultiplier tube (PMT) Hamamatsu H11526-20-NF. The optical system to collect the photons is composed of two lenses of focal lenses $f_1$=150mm and $f_2$=175 mm, with a solid angle $\Omega_{PMT}$.

The optical power collected by the PMT is proportional to the atom number $N_{at}$:

\begin{equation}
    P_{PMT} = \frac{hc}{\lambda} N_{at} \gamma_{diff} \frac{\Omega_{PMT}}{4\pi}
\end{equation}

where $\lambda=780$ nm is the optical wavelength and $c$ the speed of light. The diffusion rate $\gamma_{diff}$ is defined by eq. \ref{eq:diffusionRate}. 

The current at the output of the PMT is proportional to $P_{PMT}$: 

\begin{equation}
    I_{PMT} = \frac{hc}{\lambda} N_{at} \gamma_{diff} \frac{\Omega_{PMT}}{4\pi} \cdot S \cdot G
\end{equation}

with $S=50$mA.W$^{-1}$ at 780 nm the sensitivity of the PMT and $G=2\cdot10^6$ the gain. % for a control voltage of 0.8~V. 

%Le courant en sortie du PMT est alors transformé en tension à l'aide d'une simple résistance de R = $1193 \Omega$. 

The atom number is then calculated from the measured current: 
\begin{equation}
    N_{at} = \frac{\lambda}{hc\gamma_{diff}} \frac{4\pi}{\Omega_{PMT}} \frac{1}{S\cdot G} \cdot I_{PMT}
\end{equation}

\section*{Trap frequencies calibration}

\begin{figure}
    \centering
    \includegraphics[scale=0.35, angle=0]{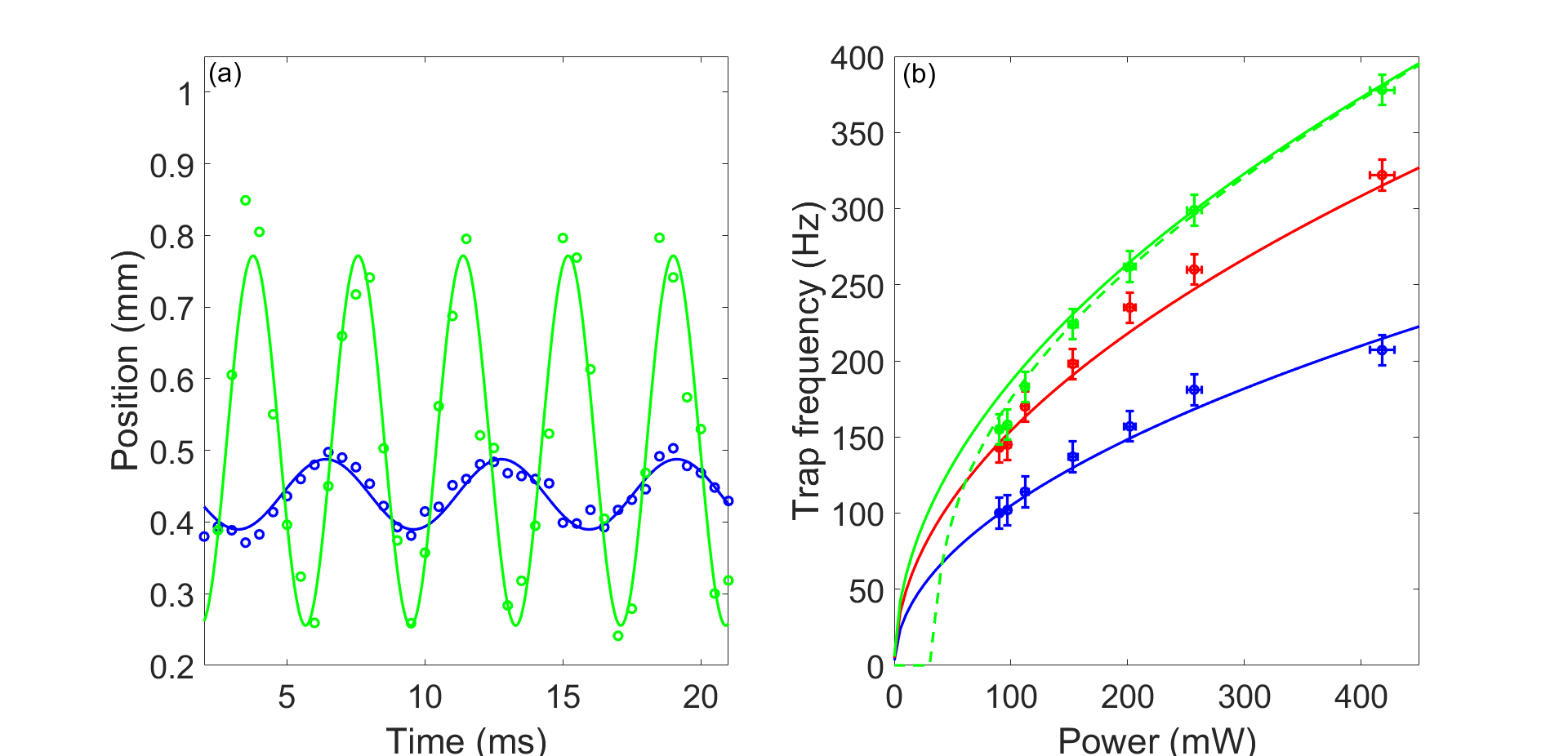}
    \caption{\textbf{Trap frequencies calibration:} (a)  Oscillations of the center of mass of the atomic cloud along the two axes X (blue) and Z (green). (b)  Trap frequencies versus the laser power in the first beam of the dipole trap (red for the Y axis). On ground, the gravity sag opens completely the trap in the Z direction corresponding to the vanishing of the trap frequency (green dashed line). In microgravity it is possible to decompress the dipole trap and reach lower trap frequencies (solid lines). The data set is consistent with a waist of the beam $\omega_0 = 42.5 \mu$m.}
    \label{fig:FrequencyTrap}
\end{figure}

The trap frequencies are calibrated in relation to the optical power in standard gravity. We excite the center of mass motion for each eigenaxis of the dipole trap by switching rapidly the position of the trap position with the AOMs or the tilt of the mirror M3 controlled by a picomotor. The frequency of the oscillation corresponds to the trap frequency (Fig \ref{fig:FrequencyTrap} (a)).  Considering the temperature in the regime of interest, the harmonic approximation is valid and we expect the same trap frequencies in microgravity. %We measured the optical power in the first laser beam of the dipole trap

As expected, the trap frequencies vary as the square root of the optical power (Fig \ref{fig:FrequencyTrap} (b)), and the data set is consistent with a laser waist $w_0=42.5 \mu$m.
On ground, gravity prevents from trapping the atoms at very low power because the dipole trap is opened by the gravity potential. The trap frequencies are extrapolated at low power and are reachable in microgravity on board the 0g aircraft. %On flight, in the microgravity phase, we reach the threshold of the Bose Einstein condensation for a mean trap depth $\omega=2\pi\cdot 59$Hz. 

\section*{DATA AVAILABILITY}
The datasets generated and/or analysed during the current study are available in the Zenodo repository 10.5281/zenodo.18633000. 

\newcommand{\newblock}{}
\bibliographystyle{unsrt}
\bibliography{biblio}

\section*{ACKNOWLEDGMENTS}
This work is supported by the French national agency CNES (Centre National d’Etudes Spatiales), the European Space Agency (ESA), and co-funded by the European Union (CARIOQA-PMP project) and PEPR (Programmes et Equipements Prioritaires de Recherche) France Relance 2030 QAFCA grant no.~ANR-22-PETQ-0005 QAFCA. For financial support, C. Pelluet and C. Metayer thank CNES, CNRS, and ESA. All authors thank Novespace for the organisation of the parabolic flights campaign.

\section*{AUTHOR CONTRIBUTIONS}
B.B. conceived the project. J.L.M., V.J., C.P. and C.M built the apparatus. J.L.M., C.M., V.J., and C.P. performed experiments, C.M. and J.L.M. carried out the data analysis. B.B. supervised the experiments and data analysis. B.B. coordinated and administrated the project. B.B., J.L.M., C.M. and V.J. wrote the manuscript. All authors discussed and reviewed the manuscript.

\section*{COMPETING INTERESTS}
The authors declare no competing interests.

\end{document}